# Terrace Aware Phylogenomic Inference from Supermatrices

**Olga Chernomor[1,2], Arndt von Haeseler[1,2] and Bui Quang Minh[1]**

[1]Center for Integrative Bioinformatics Vienna, Max F. Perutz Laboratories, University of Vienna, Medical University of Vienna, A-1030 Vienna, Austria.

[2]Bioinformatics and Computational Biology, Faculty of Computer Science, University of Vienna, A-1090 Vienna, Austria.

**Corresponding author:**

Bui Quang Minh

Center for Integrative Bioinformatics Vienna (CIBIV)

Max F. Perutz Laboratories

Dr. Bohrgasse 9

A-1030 Vienna, Austria

Tel: 0043 1 4277 24007

Fax: 0043 1 4277 24098

Email: minh.bui@univie.ac.at

**Abbreviations:** ML – Maximum Likelihood, PTA - phylogenetic terrace aware, NNI - nearest neighbor interchange, AA – amino acid.

**Running head:** Terrace aware phylogenomic inference from supermatrices

## Abstract

One approach in phylogenomics to infer the tree of life is based on concatenated multiple sequence alignments from many genes. Unfortunately, the resulting so-called supermatrix is usually sparse, that is, not every gene sequence is available for all species in the supermatrix. Due to the missing sequence information a phylogenetic inference, assuming that each gene evolves with its own substitution model, suffers from phylogenetic terraces on which many phylogenetic trees show the same likelihood. Here, we propose a phylogenetic terrace aware (PTA) data structure for efficient supermatrix based tree inference under partition models. PTA avoids likelihood computations for trees belonging to the same terrace. PTA is implemented in the IQ-TREE software, and leads to an 1.7 to 6-fold speedup for real data sets compared with a naïve implementation. Speedups are independent on terrace sizes but correlate with the amount of missing data. Thus, the PTA data structure is well suited for phylogenomic analyses. IQ-TREE source codes, binaries and documentation are freely available at http://www.cibiv.at/software/iqtree.

## Introduction

The gigantic amount of sequence data generated by next generation sequencing technologies has popularized the field of phylogenomics (Eisen 1998; Delsuc, Brinkmann & Philippe 2005; Kumar *et al.* 2012). Here, one aims to infer the tree of life from multiple genes, loci, or even whole genomes, which provide enough phylogenetic information to resolve difficult branching orders (e.g., Bininda-Emonds, Gittleman & Purvis 1999; Rokas *et al.* 2003; Dunn *et al.* 2008; Meusemann *et al.* 2010).

Phylogenomic inference methods are typically categorized into supertree and supermatrix methods (De Queiroz, Donoghue & Kim 1995; Sanderson, Purvis & Henze 1998; Bininda-Emonds, Gittleman & Steel 2002; Delsuc, Brinkmann & Philippe 2005; Kupczok, Schmidt & von Haeseler 2010). Supertree methods combine inferred gene trees into one "supertree". Supermatrix refers to the concatenation of multiple sequence alignments from different genes. Typically sequence information is not available for each gene in each species. Thus, the supermatrix can contain missing data. Traditional phylogenetic methods such as maximum likelihood (ML; Felsenstein 1981) are then used to reconstruct the species tree from the concatenated alignment (by and large ignoring the missing data). Moreover, ML implementations typically assume one substitution model for the whole alignment, which becomes problematic in a phylogenomic context because genes evolve differently. For example, heterotachy (Lopez, Casane & Philippe 2002) is prevalent in real data, where evolutionary rates vary across tree branches. Failure to account for heterotachy might cause systematic errors for all well-known reconstruction methods (Kolaczkowski & Thornton 2004; Philippe *et al.* 2005).

One natural way to account for different evolutionary scenarios is a partition model (Yang 1996) that allows genes to have their own substitution models. Three types of partition models (joint, proportional, or separate branch lengths per gene) are implemented in many ML software packages (Table 1). The model with separate branch lengths per gene, called full partition model, is the most general one and it accommodates heterotachy. However, due to missing data a full partition model may lead to what has been coined phylogenetic terrace (Sanderson, McMahon & Steel 2011), where different tree topologies have the same likelihood. Large phylogenetic terraces may hamper tree search algorithms in exploring the tree space. Even for a fixed tree topology, the full partition model may imply undefined branch lengths for branches of the gene trees corresponding to the missing data. In that sense, the original implementation in PAML (Yang 2007) did not cope adequately with missing data. Recently, Stamatakis and Alachiotis (2010) derived a so-called mesh data structure in RAxML to properly handle this issue.

Here, we introduce a phylogenetic terrace-aware (PTA) data structure for efficient tree inference with partition models in the presence of missing data. To reduce computation time PTA exploits phylogenetic terraces. PTA links branches of the species tree to the different gene trees. This mapping enables an easy topological synchronization between the species tree and gene trees after topological rearrangement of the species tree. Moreover, PTA requires a negligible computational overhead and works for the three partition models. It is therefore general and can be readily incorporated in existing ML software packages. We implemented the PTA data structure in the IQ-TREE software (Nguyen *et al.* 2014).We show that the PTA implementation substantially speeds up the tree search under partition models. Thus, the PTA data structure is useful in phylogenomic analyses.

## Methods

### *Full Partition Models*

Let $k$ denote the number of genes, loci, or codon positions of protein-coding DNA in a supermatrix. In the following we use "gene" to generally refer to any subset of genomic positions. Denote by $Y_1, Y_2, \dots, Y_k$ the species sets for the $k$ genes and $X = Y_1 \cup Y_2 \cup \dots \cup Y_k$ the set of all species. $D_1, D_2, \dots, D_k$ denote the corresponding alignments and $D$ is the concatenated alignment (supermatrix) of $D_1, D_2, \dots, D_k$. Stretches of unknown characters are added to $D$ if a species has no sequence for some gene (i.e., when $Y_i \neq X$). For a species-tree $T$, let $T|Y_i$ be the subtree of $T$ restricted to the species set $Y_i$ (see Figure 1).

In the partition model (Yang 1996) each gene $i$ evolves under its substitution model $M_i$. Moreover, a full partition model allows each gene tree $T|Y_i$ to have its own set of branch lengths $\lambda_i$ representing the number of substitutions per site. Thus, $M_{full} = \{\lambda_1, \dots, \lambda_k, M_1, \dots, M_k\}$ denotes the full partition model. The log-likelihood of a species tree $T$ is then the sum of gene tree log-likelihoods:

$$\ell(T, M_{full}|D) = \sum_{i=1}^{k} \ell(T|Y_i, \lambda_i, M_i|D_i). \quad (1)$$

Eq. (1) implies that the species tree $T$ has in fact no defined branch lengths. However, one can display the branch lengths for $T$ as the weighted average of the $\lambda_i$'s, for example.

Phylogenetic terraces can be described as follows: If two different trees $T$ and $T'$ induce the same set of gene trees (i.e., $T|Y_i = T'|Y_i,\ \forall\, i = 1, \dots, k$), then $T$ and $T'$ will have the same likelihood according to (1) and they belong to a phylogenetic terrace. The number of species trees on a phylogenetic terrace might be extremely large depending on the overlap between $Y_1, Y_2, \dots, Y_k$. For example, given two gene trees $T|Y_1$ and $T|Y_2$ in Figure 1, tree $T$ in Figure 1 belongs to a terrace of 13 trees (Sanderson, McMahon & Steel 2011). We utilized the fact that species trees belong to the same phylogenetic terrace to accelerate tree search algorithms.

***Phylogenetic Terrace Aware Data Structure***

Let $E$ denote the set of all branches of $T$ and $E_i$ the branch set of $T|Y_i$. We represent each branch $e \in E$ by its induced bipartition (or split) $e = A|B$, where $A$ and $B$ are disjoint complementary subsets of the leaf set $X$. For every gene $i$ we introduce the map

$$f_i: E \to E_i \cup \{\varepsilon\},$$

$$f_i(e) = \begin{cases} A \cap Y_i | B \cap Y_i, & A \cap Y_i \neq \emptyset \text{ and } B \cap Y_i \neq \emptyset \\ \varepsilon, & \text{otherwise.} \end{cases} \quad (2)$$

In supertree terminology, $f_i$ is the map from supersplits in $T$ to subsplits (or partial splits) in $T|Y_i$ (Semple & Steel 2003; chap. 6). Every supersplit has no or exactly one corresponding subsplit, whereas a subsplit has one or more corresponding supersplits. If a supersplit has no corresponding subsplit, we map it to $\varepsilon$. For example, in Figure 1 all supersplits from $T$ are mapped to corresponding subsplits on $T|Y_1$ (see red arrows), except the two supersplits ($\{2\}|\{1,3,4,5,6\}$ and $\{3\}|\{1,2,4,5,6\}$), thus $f_1(\{2\}|\{1,3,4,5,6\}) = \varepsilon$, $f_1(\{3\}|\{1,2,4,5,6\}) = \varepsilon$.

The collection of all maps $F = \{f_1, \dots, f_k\}$ together with the trees $\{T, T|Y_1, \dots, T|Y_k\}$ forms the data structure for partition model analyses. In the following we show how to build $F$ in $O(nk)$ time ($n = |X|$) and how to recompute $F$ in $O(k)$ time if the nearest neighbor interchange (NNI; Robinson 1971) is ap-plied to $T$. With $F$ one detects in constant time whether an NNI on $T$ will change the topology of each subtree $T|Y_i$.

### *Dynamic Programming Algorithm for Building F*

We now describe an algorithm to build $F$ in linear time for unrooted bifurcating trees using a post-order tree traversal. It first assigns $f_i$ for external branches and then proceeds towards internal branches of the tree once the two neighboring branches have already been processed. More specifically, let $e = \{x\}|X\backslash\{x\} \in E$ be an external branch then

$$f_i(e) = \begin{cases} \varepsilon, & x \notin Y_i \\ \{x\}|Y_i\backslash\{x\}, & x \in Y_i \end{cases} \quad (3)$$

Obviously, Eq. (3) follows directly from Eq. (2). Now for an internal branch $e$ where its adjacent branches $e_1, e_2$ have already been processed, we assign $f_i(e)$ as follows:

$$f_i(e) = \begin{cases} \varepsilon, & f_i(e_1) = f_i(e_2) \\ f_i(e_1), & f_i(e_1) \neq \varepsilon \text{ and } f_i(e_2) = \varepsilon \\ f_i(e_2), & f_i(e_1) = \varepsilon \text{ and } f_i(e_2) \neq \varepsilon \\ \text{branch adjacent to } f_i(e_1), f_i(e_2), & \text{otherwise.} \end{cases} \quad (4)$$

**Proof for the correctness of eq. (4).** Figure 2a illustrates $T$ around $e_1, e_2, e$, where $A, B, C$ are the three corresponding species sets. We note that $e = (A \cup B)|C$, $e_1 = A|(B \cup C)$, $e_2 = B|(A \cup C)$.
From eq. (2) it follows, that

$$f_i(e) = \begin{cases} (A \cup B) \cap Y_i | C \cap Y_i, & (A \cup B) \cap Y_i \neq \emptyset \text{ and } C \cap Y_i \neq \emptyset \\ \varepsilon, & \text{otherwise} \end{cases}, \quad (5)$$

$$f_i(e_1) = \begin{cases} A \cap Y_i | (B \cup C) \cap Y_i, & A \cap Y_i \neq \emptyset \text{ and } (B \cup C) \cap Y_i \neq \emptyset \\ \varepsilon, & \text{otherwise} \end{cases}, \quad (6)$$

$$f_i(e_2) = \begin{cases} B \cap Y_i | (A \cup C) \cap Y_i, & B \cap Y_i \neq \emptyset \text{ and } (A \cup C) \cap Y_i \neq \emptyset \\ \varepsilon, & \text{otherwise} \end{cases}. \quad (7)$$

We now consider the four cases from eq. (4):

1) $f_i(e_1) = f_i(e_2)$. If they are equal to $\varepsilon$, then from (6) and (7) it follows that at least two of the three intersections $A \cap Y_i$, $B \cap Y_i$ and $C \cap Y_i$ are empty. Therefore from Eq. (5) we have $f_i(e) = \varepsilon$. Otherwise if $f_i(e_1) = f_i(e_2) \neq \varepsilon$, then we have $A \cap Y_i = (A \cup C) \cap Y_i$ and $B \cap Y_i = (B \cup C) \cap Y_i$, from which it follows that $C \cap Y_i = \emptyset$ and thus $f_i(e) = \varepsilon$.

2) $f_i(e_1) \neq \varepsilon$ and $f_i(e_2) = \varepsilon$. From $f_i(e_1) \neq \varepsilon$ it follows that $A \cap Y_i \neq \emptyset$ and $(B \cup C) \cap Y_i \neq \emptyset$, while from $f_i(e_2) = \varepsilon$, $B \cap Y_i = \emptyset$ or $(A \cup C) \cap Y_i = \emptyset$. Since $A \cap Y_i \neq \emptyset$ then $(A \cup C) \cap Y_i \neq \emptyset$, and therefore $B \cap Y_i = \emptyset$ and $C \cap Y_i \neq \emptyset$. Since sets $A \cap Y_i$ and $C \cap Y_i$ are not empty while $B \cap Y_i$ is, then

$$f_i(e) = (A \cup B) \cap Y_i | C \cap Y_i = A \cap Y_i | (B \cup C) \cap Y_i = f_i(e_1).$$

3) $f_i(e_1) = \varepsilon$ and $f_i(e_2) \neq \varepsilon$. Symmetrically to condition 2 above, we have $f_i(e) = f_i(e_2)$.

4) $f_i(e_1) \neq f_i(e_2) \neq \varepsilon$. From $f_i(e_1) \neq \varepsilon$ we have that $A \cap Y_i \neq \emptyset$, from $f_i(e_2) \neq \varepsilon$ it follows that $B \cap Y_i \neq \emptyset$, and since $f_i(e_1) \neq f_i(e_2)$ then $C \cap Y_i \neq \emptyset$. Therefore, $f_i(e) \neq \varepsilon$ is a branch on subtree $T|Y_i$ incident to $f_i(e_1)$ and $f_i(e_2)$ (Figure 2b).

Thus, eq. (4) is correct.■

***Quick Nearest Neighbor Interchange Using F***

We start with the following observation:

**Observation 1.** Let $e \in E$ be an internal branch and $e_1, e_2, e_3, e_4 \in E$ the four branches adjacent to $e$, then an NNI around $e$ will change the tree topology of $T|Y_i$ iff all $f_i(e), f_i(e_1), f_i(e_2), f_i(e_3), f_i(e_4) \neq \varepsilon$.

**Proof.** *Sufficiency*. W.l.o.g. we assume that the subtrees belonging to $e_1$ and $e_3$ are exchanged via NNI. Let $A, B, C, D$ denote the corresponding species sets leading from $e_1, e_2, e_3, e_4$, respectively (Figure 2). Therefore, $A \cap Y_i, B \cap Y_i, C \cap Y_i, D \cap Y_i$ are the species sets represented by sequences from gene $i$. If one of these four species sets were empty, then the NNI operation on $T$ would not change the topology of $T|Y_i$. Thus $A \cap Y_i \neq \emptyset, B \cap Y_i \neq \emptyset, C \cap Y_i \neq \emptyset, D \cap Y_i \neq \emptyset$. Then $f_i(e_1) = A \cap Y_i | (B \cup C \cup D) \cap Y_i \neq \varepsilon$. Similarly one computes $f_i(.)$ for the other branches. This proves the sufficiency.

*Necessity*. Let $f_i(e), f_i(e_1), f_i(e_2), f_i(e_3), f_i(e_4) \neq \varepsilon$. W.l.o.g. we assume that $e_1, e_2$ are two adjacent branches. Because $f_i(e), f_i(e_1), f_i(e_2) \neq \epsilon$, it follows from Eq. (4) that these three branches are adjacent in $T|Y_i$. Similarly, $f_i(e), f_i(e_3), f_i(e_4)$ are also adjacent branches. That means, the four branches $f_i(e_1), f_i(e_2), f_i(e_3), f_i(e_4)$ are incident to $f_i(e)$. Therefore, an NNI on $e$ of $T$ by swapping $e_1$ and $e_3$ corresponds to an NNI on $T|Y_i$ by swapping $f_i(e_1)$ and $f_i(e_3)$. ■

With Observation 1 we identify in constant time if an NNI on $T$ changes the topology of $T|Y_i$. Let $T_{NNI}$ denote the tree after swapping $e_1$ and $e_3$. One then updates the gene tree $T|Y_i$ using two rules:

1. If $f_i(e), f_i(e_1), f_i(e_2), f_i(e_3), f_i(e_4) \neq \varepsilon$, then $T_{NNI}|Y_i$ will result from $T|Y_i$ by swapping $f_i(e_1), f_i(e_3)$.

2. Otherwise, we have $T_{NNI}|Y_i = T|Y_i$. Thus, we keep the tree topology $T|Y_i$ and only have to update $f_i(e)$ according to (4).

Obviously these rules enable a linear-time update for all induced gene trees to be synchronized with $T_{NNI}$. As an illustration, for the tree $T$ shown in Figure 1 the two NNIs around the branch $\{4,6\}|\{1,2,3,5\}$ will change the topology of $T|Y_1$ but do not influence the topology of $T|Y_2$.

Observation 1 also reduces the log-likelihood computation of $T_{NNI}$ (Eq. 1) as the log-likelihood of the gene tree only changes when $T_{NNI}|Y_i \neq T|Y_i$. Especially, when $T$ and $T_{NNI}$ belong to the same phylogenetic terrace, then no re-computation of the log-likelihood is needed.

### *Partition Model With Proportional Branch Lengths*

The proportional branch length model assumes one set of branch lengths $\lambda$ for the species tree $T$ and rescales the gene trees with specific positive, non-zero rates $r_1, r_2, \dots, r_k$ such that the average rate is 1 (i.e., $\sum_{i=1}^{k} r_i = k$). We denote such a model $M_{prop} = \{\lambda, r_1, \dots, r_k, M_1, \dots, M_k\}$, which is a special case of $M_{full}$. The branch length $\lambda_i$ for each gene tree is determined by:

$$\lambda_i(e') = r_i \times \sum_{e \in E: f_i(e) = e'} \lambda(e), \quad \forall e' \in E_i. \qquad (4)$$

All parameters of $M_{prop}$ are estimated by maximizing the log-likelihood function (1) under the constraint (4). Similarly to $M_{full}$, model parameters $M_i$ and gene rates $r_i$ of $M_{prop}$ are optimized separately for each gene tree. However, while $M_{full}$ optimizes branch lengths per gene tree, $M_{prop}$ optimizes branch lengths of the species tree. To speed up optimization of a branch $e \in E$ using eq. (1), one only has to sum the log-likelihoods over those genes $i$ where $f_i(e) \neq \varepsilon$.

### *Time complexity*

We now assess the time complexity of the PTA data structure. First, for a fixed tree the computation of $F$ needs $O(nk)$ time, where $n$ is the number of species and $k$ the number of genes, because each map $f_i$ is constructed by one tree traversal in $O(n)$ time (see Section 2.3). Second, when changing the topology of $T$ by one NNI, $F$ is recomputed in $O(k)$ time because each $f_i$ is updated in constant time (see Section 2.4). Therefore, the total time needed to maintain $F$ during tree search is $O(nk)$. This extra computation is negligible compared with the expensive likelihood computations.

### *Benchmark Setup*

We provide IQ-TREE$_{PTA}$, the version of the IQ-TREE software that implemented the PTA data structure. To assess the performance of IQ-TREE$_{PTA}$ against a naïve version, we analyzed five DNA and three AA alignments (Table 2). Alignments 1, 2, 4 were studied by Stamatakis and Alachiotis (2010), alignment 3 is from Bouchenak-Khelladi *et al.* (2008), alignment 5 is from Pyron *et al.* (2011), and alignments 6, 7, 8 are from Dell'Ampio *et al.* (2014). For alignments 3, 4 and 5 it was shown that the highest ML tree found for each alignment belongs to a large terrace, comprising of up to 236 million, more than 1 billion and 11,025 species trees with equal likelihoods respectively.

We applied the full partition model assuming a GTR+$\Gamma$ (Lanave *et al.* 1984; Yang 1994) and LG+$\Gamma$ (Yang 1994; Le & Gascuel 2008) models for all genes in the DNA and the AA alignments, respectively. We estimated the substitution parameters for each gene separately. For each alignment we performed 10 independent IQ-TREE$_{PTA}$ tree reconstruction runs. We then recorded the average CPU times until the best tree was found. All computations were carried out on a homogeneous cluster of 3.3 GHz computers.

## Results

### *Computational Efficiency*

To assess the computing efficiency of IQ-TREE$_{PTA}$ we analyzed the eight alignments (Table 2). The number of genes ranges from 3 to 51 and the amount of missing data in the supermatrix varies from 28% (alignment 1) and 73% (alignment 4).

Table 2 shows the percentage of NNIs swaps for each alignment that did not change the tree structure of the gene trees (see Observation 1; Methods). The fraction ranges from 39% to 83%, where the saving of operation correlates with the amount of missing data. Table 2 also shows the average CPU time for each alignment. Because the NNI search is the most time-consuming component of the tree search, the naïve IQ-TREE version runs much longer (see IQ-TREE$_{Naive}$ column). In fact the computational speed up of the terrace aware data structure ranges from 1.7 (alignment 1) to 6.0 (alignment 4). For alignment 4 IQ-TREE$_{PTA}$ needed approximately 4 hours, whereas IQ-TREE$_{Naive}$ ran nearly one day, clearly showing that the PTA data structure makes efficient use of the large amount of missing data (73%) in this alignment.

### *Phylogenetic Terrace Sizes*

We computed the phylogenetic terrace sizes for the best ML trees found, adapting the script from Sanderson, McMahon and Steel (2011). For alignments 1, 2, 6, 7 and 8 the terrace sizes varied between 1 and 3 across 10 runs per alignment. Alignments 5, 3, and 4 show moderate, large, and huge terrace sizes, respectively (Table 2). Notably, all 10 ML trees inferred from alignment 4 have more than 1 billion species trees showing the same likelihood. Alignment 4 is certainly not suitable for phylogenomic analysis assuming the full partition model.

We note that for the five alignments with small terrace sizes, IQ-TREE$_{PTA}$ still achieved speedup factors of 1.6 to 1.8 compared with IQ-TREE$_{Naive}$. Therefore, terrace sizes are not indicative of computational efficiency.

### *Comparisons with RAxML*

We also performed 10 independent RAxML 8.0.2 (Stamatakis 2014) partition model analyses (-M option) for each alignment. Table 2 shows the highest (out of 10) likelihoods for each alignment computed by IQ-TREE$_{PTA}$ and RAxML, respectively. The IQ-TREE$_{PTA}$ runs yielded higher likelihoods than RAxML for alignments 2 and 4. RAxML was better for alignments 3 and 5. For the remaining four alignments the log-likelihood differences were less than 10 units. Therefore, neither program performs best for all alignments.

Table 2 also shows the average CPU times needed by RAxML. IQ-TREE$_{Naive}$ is much slower than RAxML except for alignment 2. However, IQ-TREE$_{PTA}$ is faster than RAxML for alignments 2 and 8 whereas RAxML is faster for alignments 3, 4 and 5. For the remaining alignments both programs need similar amount of CPU time.

## Conclusions

We have presented a very efficient PTA data structure by exploiting the property of phylogenetic terraces, which speeds up phylogenomic inference under partition models. We showed that PTA awareness accelerates NNI searches. Moreover, our implementation (IQ-TREE$_{PTA}$) not only allows individual substitution models per gene but also joint, proportional, or separate branch lengths and mixed DNA/AA partitions within one partition model analysis. Apart from TreeFinder, IQ-TREE$_{PTA}$ is more versatile than other ML software (see Table 1).

We briefly compared IQ-TREE$_{PTA}$ with RAxML in terms of computing times. From the small number of test alignments, we conclude that IQ-TREE$_{PTA}$ is as fast as RAxML in most cases. However, a more systematic evaluation is necessary.

Finally, we plan to extend the PTA data structure for other tree rearrangement operations such as subtree pruning and regrafting. We also plan to implement the heuristic searches for the best-fit partitioning scheme (Lanfear *et al.* 2012), thus allowing users to perform the partition model selection and ML inference within one single run.

## Acknowledgements

*Funding*: This work was supported by the Austrian Science Fund - FWF (grant number I760-B17) to B.Q.M. and A.v.H., and the University of Vienna (Initiativkolleg I059-N) to O.C and A.v.H.

**Table 1.** Availability of partition models in ML tree search software.

| Software | Joint | Proportional | Separate |
|---|---|---|---|
| MetaPIGA (Helaers & Milinkovitch 2010) | x | x | |
| PhyML (Guindon *et al.* 2010) | | | |
| GARLI (Zwickl 2006) | x | x | |
| RAxML (Stamatakis 2006) | x | | x |
| TreeFinder (Jobb, von Haeseler & Strimmer 2004) | x | x | x |
| IQ-TREE (Nguyen *et al.* 2014) | x | x | x |

**Table 2.** Benchmark alignments and analysis results.

| No. | Type | No. species | No. genes | No. sites | Missing data | %NNIs avoided | Average CPU time | | | Terrace size | Tree log-likelihood | |
|---|---|---|---|---|---|---|---|---|---|---|---|---|
| | | | | | | | IQ-TREE$_{Naive}$ | IQ-TREE$_{PTA}$ | RAxML | | IQ-TREE$_{PTA}$ | RAxML |
| 1 | DNA | 59 | 8 | 6,951 | 28% | 39% | 9m:06s | 5m:28s | 0h:3m:13s | 1 - 3 | -50,391 | -50,392 |
| 2 | DNA | 128 | 34 | 29,198 | 30% | 39% | 1h:28m:35s | 0h:54m:02s | 1h:41m:57s | 1 | -779,116 | -779,167 |
| 3 | DNA | 298 | 3 | 5,074 | 34% | 43% | 3h:21m:48s | 1h:55m:01s | 0h:43m:57s | 429 K - 236 Mio. | -69,659 | -69,638 |
| 4 | DNA | 404 | 11 | 13,158 | 73% | 83% | 23h:8m:37s | 3h:56m:03s | 2h:20m:3s | $>10^9$ | -150,751 | -150,810 |
| 5 | DNA | 767 | 5 | 5,714 | 59% | 53% | 23h:44m:37s | 11h:23m:49s | 3h:32m:7s | 175 - 11,025 | -369,115 | -368,904 |
| 6 | AA | 69 | 31 | 8,546 | 35% | 42% | 3h:26m:34s | 2h:01m:52s | 1h:50m:23s | 1 | -179,753 | -179,752 |
| 7 | AA | 70 | 35 | 11,789 | 34% | 42% | 5h:15m:12s | 3h:05m:58s | 3h:35m:6s | 1 | -249,470 | -249,464 |
| 8 | AA | 72 | 51 | 12,548 | 35% | 44% | 4h:40m:12s | 2h:34m:06s | 3h:40m:40s | 1 - 3 | -329,024 | -329,024 |

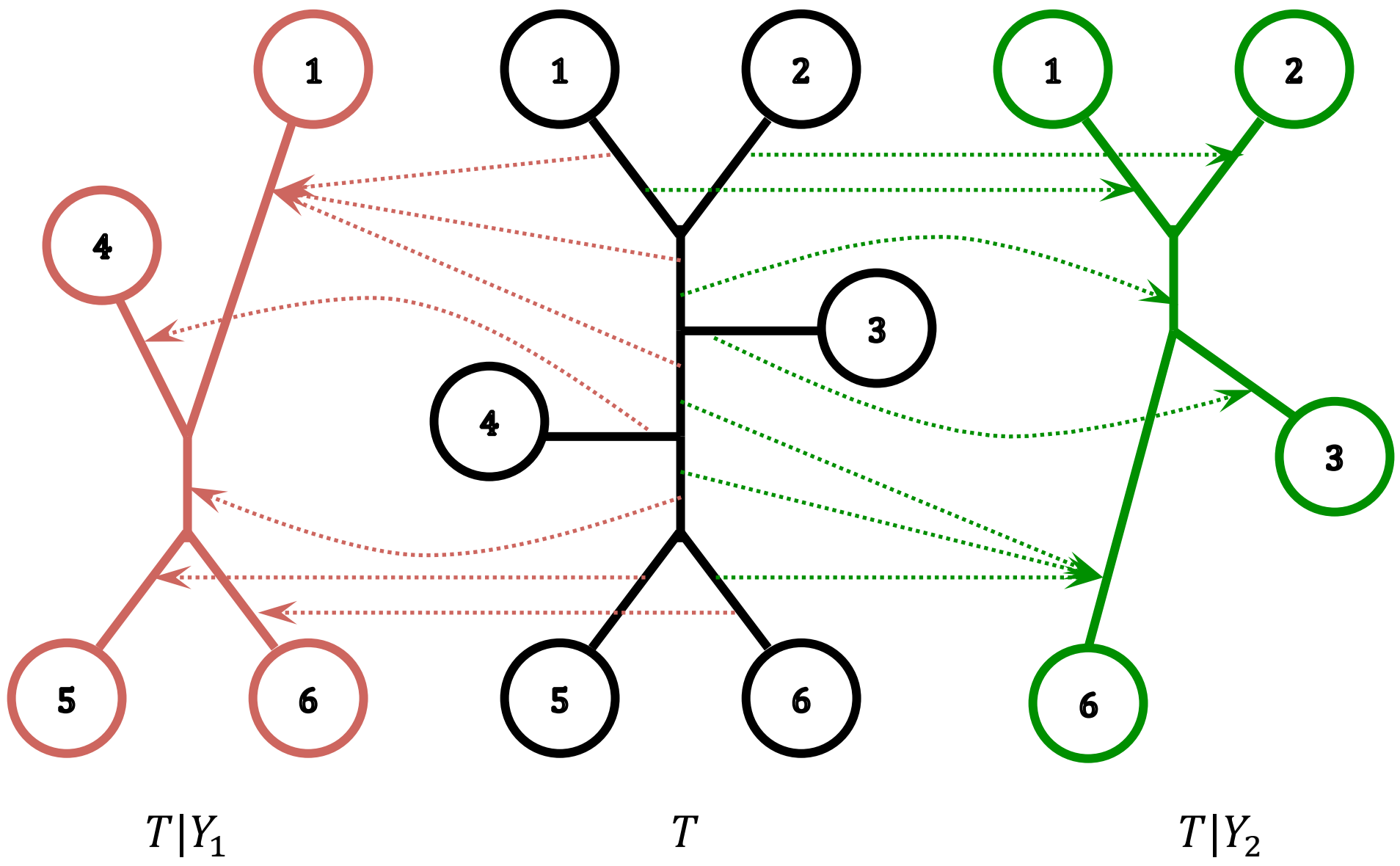

**Fig. 1.** A species tree $T$ with species $X = \{1,2,3,4,5,6\}$ and two induced gene trees $T|Y_1$ and $T|Y_2$ with species sets $Y_1 = \{1,4,5,6\}$ and $Y_2 = \{1,2,3,6\}$.

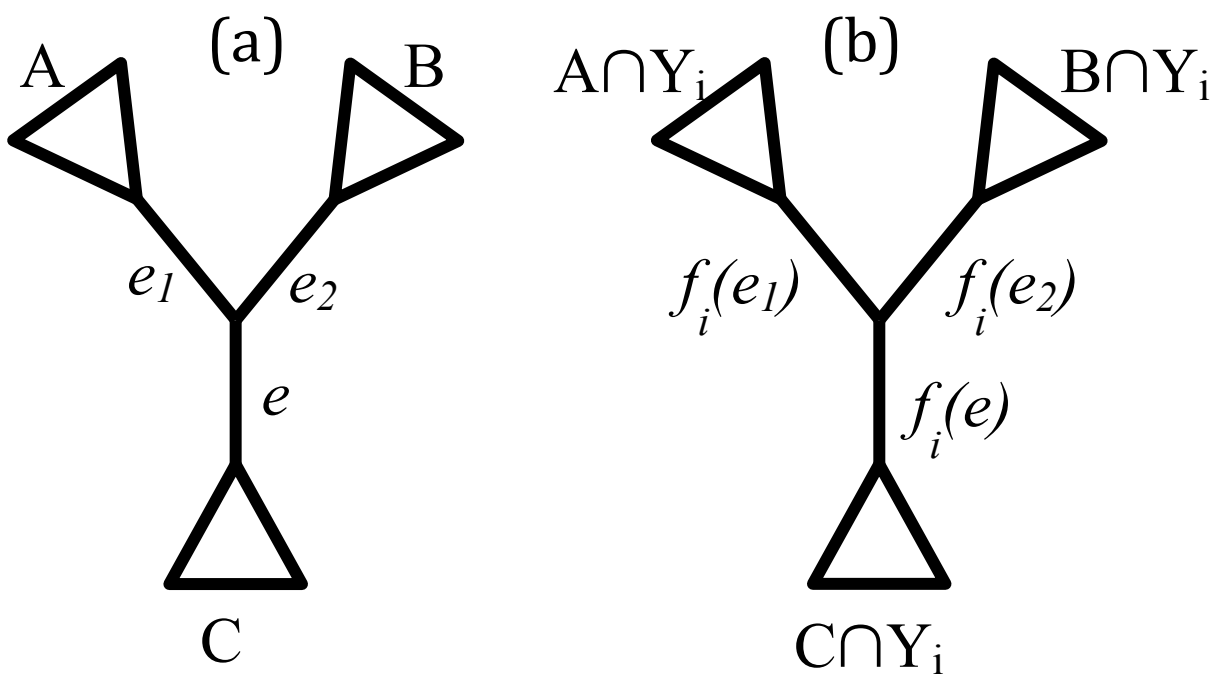

**Fig. 2.** Proof for the correctness of eq. (4). (a) Three adjacent branches on species tree $T$ and (b) their corresponding branches on gene tree $T|Y_i$.

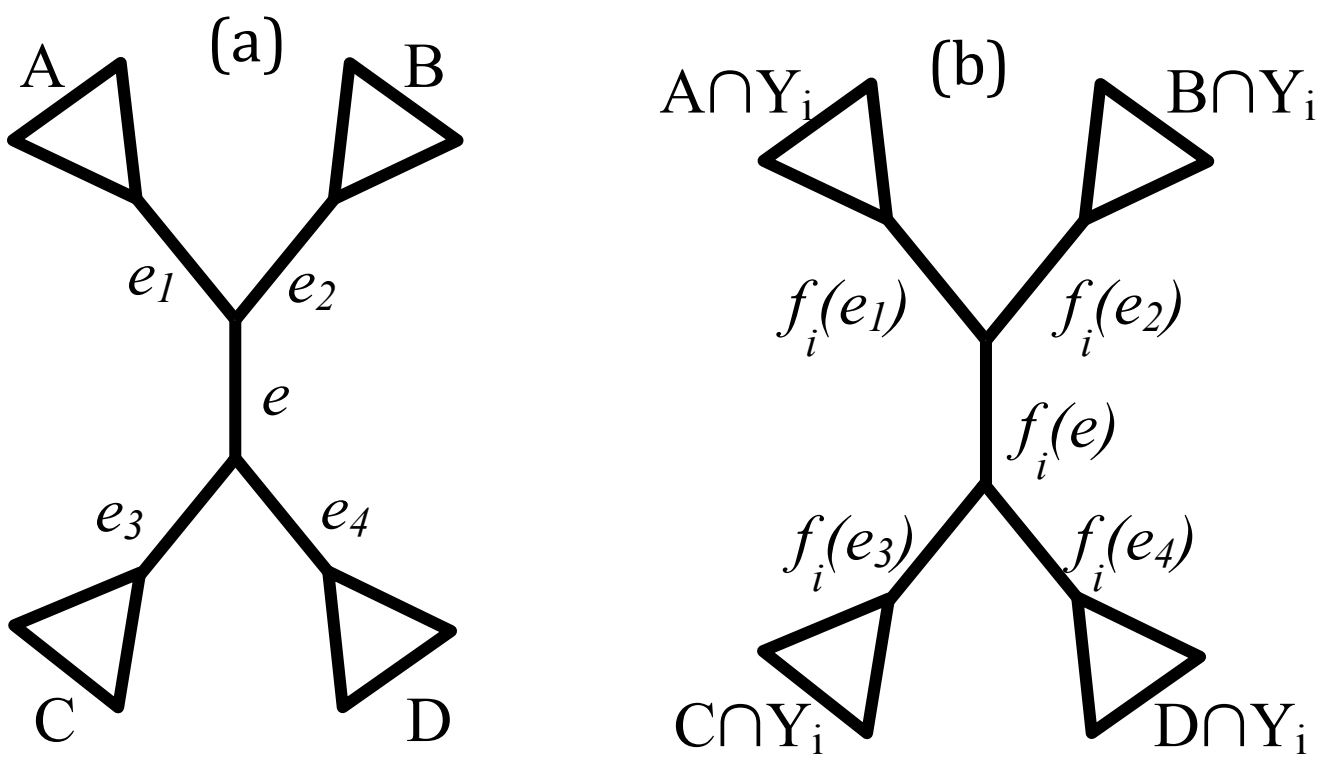

**Fig. 3.** Illustration for the proof of observation 1. (a) Species tree $T$ around internal branch $e$ and (b) Corresponding branches on gene tree $T|Y_i$.